Time-Resolved Ultrafast Transient Polarization Spectroscopy to Investigate Nonlinear Processes and Dynamics in Electronically Excited Molecules on the Femtosecond Time Scale

Richard Thurston[1], Matthew M. Brister[1], Ali Belkacem[1], Thorsten Weber[1], Niranjan Shivaram[1,2] and Daniel S. Slaughter[1]*

1. Chemical Sciences Division, Lawrence Berkeley National Laboratory, Berkeley, California, USA 94720

Abstract

We report a novel experimental technique to investigate ultrafast dynamics in photoexcited molecules by probing the third-order nonlinear optical susceptibility. A non-colinear 3-pulse scheme is developed to probe the ultrafast dynamics of excited electronic states using the optical Kerr effect by time-resolved polarization spectroscopy. Optical homodyne and optical heterodyne detection are demonstrated to measure the 3rd-order nonlinear optical response for the $S_1$ excited state of liquid nitrobenzene, which is populated by 2-photon absorption of a 780 nm 40 fs excitation pulse.

Introduction

Recent developments in high intensity and high repetition rate femtosecond-pulsed laser systems have enabled the development of nonlinear pump-probe spectroscopic techniques with high temporal resolution and high sensitivity to the dynamics of excited molecules. Some of these techniques include time-resolved photoelectron spectroscopy[1], transient absorption spectroscopy,[2] fluorescence up-conversion,[3] pump degenerate four-wave mixing,[4] and third harmonic generation.[5] For many molecular systems, the 3rd-order optical susceptibility $\chi^{(3)}$ is the lowest order nonlinear optical response[6] and many techniques have been developed that use $\chi^{(3)}$ as a sensitive probe of electronic structure and excited state dynamics.[7–10] With additional parameters such as independent frequencies of two or more probe pulses, multiple time delays, and phase-matching conditions, nonlinear spectroscopy can be highly differential and exquisitely sensitive to dynamics such as electron-hole coherence, or populations of specific electronic states. However, such differential experimental schemes are still rather scarce, and the investigation of excited electronic states could be further advanced to offer a higher level of detail.

Molecules in excited electronic states may exhibit a large second hyperpolarizability, which is the molecular property contributing to $\chi^{(3)}$, if strong dipole coupling exists between the excited state and virtual intermediate states, or if a high density of nearby real and virtual electronic states is present.[5,7,11,12] Techniques that utilize this enhancement of $\chi^{(3)}$ to study dynamics of electronically excited systems include third harmonic generation, which has been applied as a probe to study the dynamics of fishnet metamaterials (multilayer nanoscale patterned solid state materials) as well as atmospheric air samples[8,13]. In these experiments the excitation pulse is followed by an intense near infrared (NIR) pulse that generates the third harmonic signal. For isotropic media, this method is sensitive to a single nondegenerate tensor element of the 3rd-order susceptibility $\chi^{(3)}_{xxxx}$.[6] Four-wave mixing techniques have also been applied to the study of excited electronic state dynamics. In pump degenerate four-wave mixing experiments,[4] excited state enhancement of $\chi^{(3)}$ has been demonstrated for a variety of systems, including retinal derivatives,[14] diphenylhexatriene,[7] flame atomized gold,[15] and doped polymer film[16] with order of magnitude or greater enhancement being reported. In each of these experiments, a resonant pump pulse populates an excited electronic state, this fraction then interacts through $\chi^{(3)}$ with a pair of degenerate pulses to produce a transient grating that scatters photons along predicable momentum vectors, producing the nonlinear signal for specific phase-matching conditions.[6]

We report a novel technique that uses one femtosecond laser pulse to excite a molecule and a pair of non-resonant femtosecond probing pulses to initiate a response to the system's 3rd-order susceptibility through a time-resolved optical Kerr effect (OKE) measurement. OKE spectroscopy is a well-established approach to probe the 3rd- order susceptibility of a medium using two noncolinear laser pulses, where the measured signal is colinear with one of the pulses.[17,18] In this measurement, one pulse induces a transient birefringence through the medium's intensity dependent refractive index, and the second pulse interacts with this birefringence resulting in a rotation of the polarization of the very same pulse.[19–22] The effective 3rd- order susceptibility contains contributions from specific tensor elements of $\chi^{(3)}$.[21,23,24] Optical homodyne measurements are sensitive to the square magnitude of $\chi^{(3)}$, while optical heterodyning provides access to further detail in the real and imaginary parts of $\chi^{(3)}$. Optical heterodyning can be performed by delivering a local oscillator pulse, derived from the probe pulse, to measure the intensity dependent refractive index as a function of time-delay between two pulses.[6,21,22]

The addition of an excitation pulse to OKE spectroscopy enables the measurement of coupled electronic and nuclear dynamics on the excited state through the evolution of the system's third order optical susceptibility. In this scheme, which we call Ultrafast Transient Polarization Spectroscopy (UTPS), the excited state 3rd-order response is measured by the OKE, to probe the effective 3rd-order susceptibility, $\chi^{(3)}_{eff}$, of the molecular system. By repeating the measurement for a series of delays relative to the excitation pump pulse, UTPS applies the ultrafast transient polarization of the probe as a measurement of $\chi^{(3)}_{eff}$, for femtosecond time-resolved investigation of molecular dynamics involving electronic excited states.

UTPS offers several advantages in measuring the nonlinear response of a molecular system. Only two pulses are required to probe the 3rd-order response of a sample, requiring fewer mirrors and optical delay lines than some other nonlinear spectroscopic techniques. Optical homodyne or heterodyne detection can be achieved with the same optical instrumentation, since a complete heterodyne measurement only requires an additional quarter-wave plate[21,23,24]. Phase matching enforces a specific polarization for the emitted signal in the direction of either probe beam, so that the system's 3$^{rd}$-order response is extracted with very low background using crossed polarizers. The colinear probe and signal allows a range of possible angles for the pump beam and for each probe beam. The probe pulses may be tuned to resonant or non-resonant wavelengths, and in the non-resonant case the 3$^{rd}$-order susceptibility is measured in a weak interaction with the medium, avoiding or reducing unwanted photoreactions in the sample.

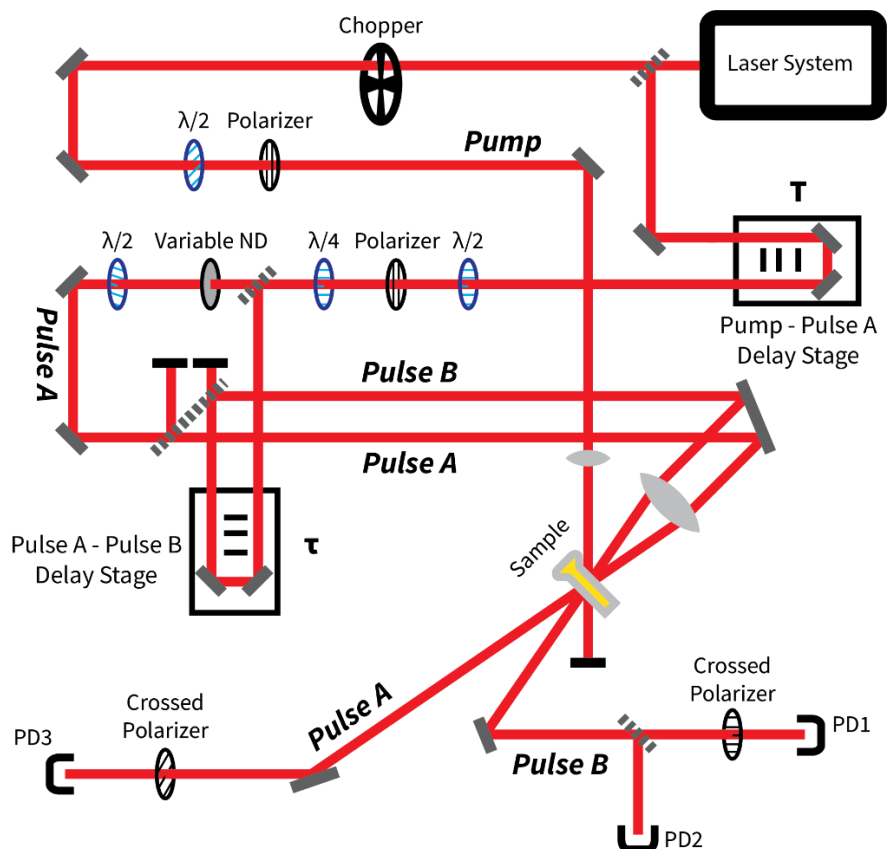

Figure 1: Schematic of the UTPS apparatus. A 1 kHz 780 nm Ti-Sapphire pulsed laser source with 35 fs pulse duration is split into 3 pulses: an excitation pump pulse, and 2 probe pulses *A* and *B*. Changes in polarization due to the transient birefringence are measured using a pair of high contrast polarizers with the output coupled onto sensitive photodiodes. All pulses are within 3º of normal incidence on the sample cuvette. The quarter-wave plate in the pulse *A* and *B* beam path allows out-of-phase heterodyne measurements to be performed by introducing a small amount of ellipticity. For homodyne and in-phase heterodyne measurements the fast axis of the quarter-wave plate is aligned parallel to the probe *B* polarization.

Methods

In the present experiments, NIR laser pulses, having a central wavelength of 780 nm, a duration of 40 fs and a repetition rate of 1 kHz, are produced with an amplified Ti:sapphire laser system. Each pulse is split into three separate pulses, an excitation pump pulse, a probe pulse *A*, and a probe pulse *B* (see Figure 1), using UV fused silica plate beamsplitters. Each pulse is focused using a 30 cm lens to intersect at the sample in a non-colinear geometry with small angles (less than 3º). Spatial and temporal overlap at the sample is achieved before and after each experiment by optimizing the sum frequency generation with each pulse pair in a beta-barium borate crystal at the sample position. Each pulse intensity is adjusted using half-wave plates and polarizers, and a variable neutral density filter for probe pulse *A*, typically to 1 x 10$^{11}$ W/cm$^2$ for the excitation pump pulse, and 3 x 10$^{10}$ W/cm$^2$ for each of the probe pulses *A* and *B*. Probe pulse *A*, with a polarization set using a half-wave plate to 45° relative to probe pulse *B*, initiates a transient birefringence in the medium, with which probe pulse *B* interacts. A pair of crossed ultra-high contrast polarizers (Meadowlark Optics, contrast >10$^6$) on the probe *B* beamline enables any changes in the probe polarization, due to the intensity dependent refractive index of the media, to be detected. Two photodiode detectors, PD3 and PD1, are employed to collect the polarization signals emitted along the direction of probe *A* and probe *B*, respectively. A third photodiode detector, PD2, measures the transmission of probe *B*. While we focus here on signals detected in the probe *B* direction (PD1), the probe *A* beamline also has crossed polarizers, and the signals detected in the probe *A* direction (PD3) are also recorded. With this arrangement, two measurements can be performed simultaneously, each having different signal polarization relative to the pump polarization. The pump pulse train is modulated with a mechanical chopper to a reference frequency of 137 Hz, for use with a lock-in amplifier (SRS SR830). This ensures that the recorded signals always include the pump pulse interaction. Specifically, the detected total signal includes the pump-probe (2-pulse) interaction with the media and the pump-*A*-*B* (3-pulse) interaction with the media. The 2-pulse signal is measured with the probe pulse *A* blocked, and the 3-pulse interaction is determined by subtracting the 2-pulse signal from the total signal.

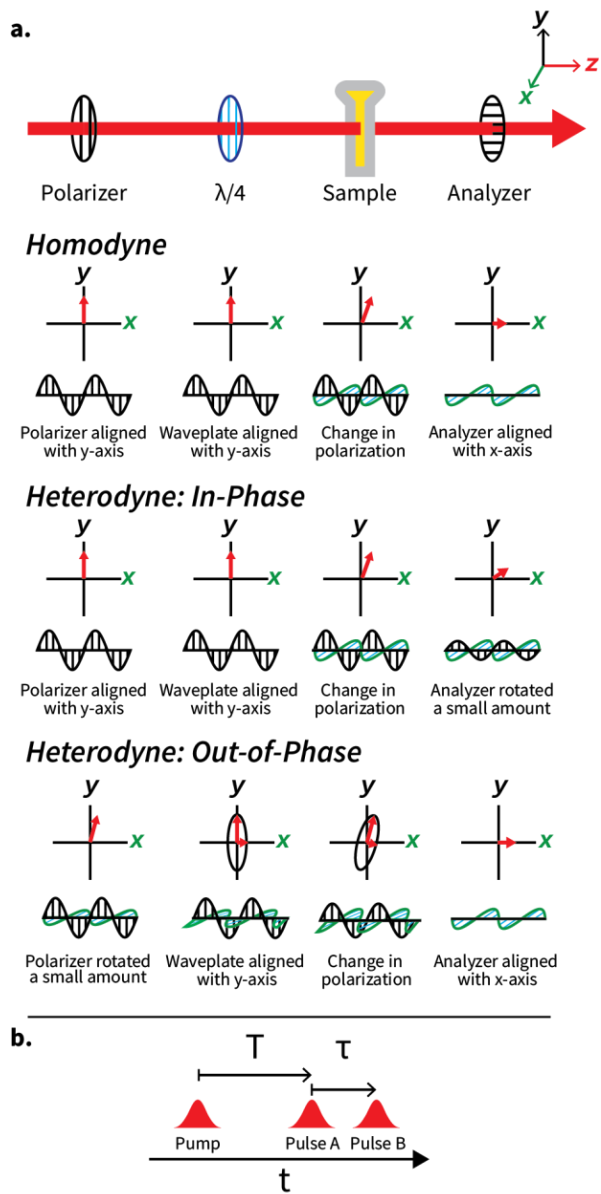

Figure 2: (a) Three different polarization schemes that are used for homodyne, in-phase heterodyne, and out-of-phase heterodyne UTPS experiments. In the homodyne case, the first polarizer aligns the measured pulse along the lab-frame y-axis and the quarter-wave plate doesn't change the polarization of input pulse. The polarization of the measured pulse is modified due to interaction with the sample and an analyzer is aligned to the x-axis to reject light with polarization along the y-axis. In the in-phase heterodyne case, the analyzer is rotated by ±ε so that a small in-phase component along the y axis is transmitted to the detector. For the heterodyne out-of-phase measurement, the first polarizer is rotated by ±ε which introduces a small off-axis component into the pulse. The quarter-wave plate introduces ellipticity in the probe pulse. Both +ε and -ε measurements are required for each heterodyne signal. (b) Time delay scheme used in the UTPS apparatus. For positive time delays, the excitation pump pulse arrives first at the sample, then the *A-B* pulse pair arrives with a time delay *T* relative to the pump pulse. The probe pulse *B* arrives with a time delay τ after the probe *A* pulse.

With the use of a quarter-wave plate after the polarizer in the probe *B* line, the apparatus can be configured to make either a homodyne or heterodyne measurement of the sample (Figure 2a). In the homodyne configuration, the probe *B* pulse is polarized in the y-direction before interacting with the sample, and the analyzer is aligned to the orthogonal x-direction. A heterodyne measurement can be made using either an in-phase, or a $\pi/2$ out-of-phase, local oscillator. For the in-phase heterodyne measurement, the input polarization of the probe *B* pulse has no ellipticity, and the analyzer is rotated by a small angle ±ε to sample a portion of the probe *B* pulse along the y-axis. This portion is the local oscillator, and it is in-phase with the signal along the x-axis. The out-of-phase heterodyne measurement involves introducing a small amount of ellipticity into the probe *B* pulse by rotating the polarizer away from the y-axis by a small angle ±ε. A quarter-wave plate (fast axis aligned to the y-direction) introduces a small ellipticity to the probe *B* polarization. This out-of-phase component of the input polarization along the direction of the analyzer is the local oscillator. Two optical delay stages (Figure 1) control the time delays, *T* and τ, between each pulse. For positive time delays, the excitation pump pulse arrives first at the sample, then the *A*-*B* pulse pair arrives with a time delay *T* relative to the pump pulse. The probe pulse *B* arrives with a time delay τ after the probe *A* pulse.

We apply the technique to measure the coupled electronic and nuclear dynamics of liquid nitrobenzene at room temperature. The nitrobenzene sample is contained in a Spectrosil quartz cuvette, having a sample optical path length of 1 mm and wall thickness of 1 mm. The intensity of the 780 nm pump pulse was optimized for 2-photon excitation to the first singlet excited state $S_1$, by measuring the transmission of the pump pulse for a broad range of intensities. The nitrobenzene sample was exchanged at regular intervals to avoid any significant concentration of photoproducts that could contribute to the measurements. The time-resolved spectrogram of each of the pump, probe *A*, and probe *B* pulses was recorded by frequency-resolved optical gating (FROG).

Results

The measured OKE signal is dependent on the effective 3rd-order optical susceptibility, $\chi_{eff}^{(3)}$, which is defined by the relative polarizations of the probe *A* and probe *B* electric fields, and the polarization of the measured field. In the present experiments $\chi^{(3)}_{eff}$ includes contributions from molecules in the ground and excited electronic states:

$$\chi^{(3)}_{eff} = \chi^{(3)}_{eff\ ground} + \chi^{(3)}_{eff\ excited} \tag{1}$$

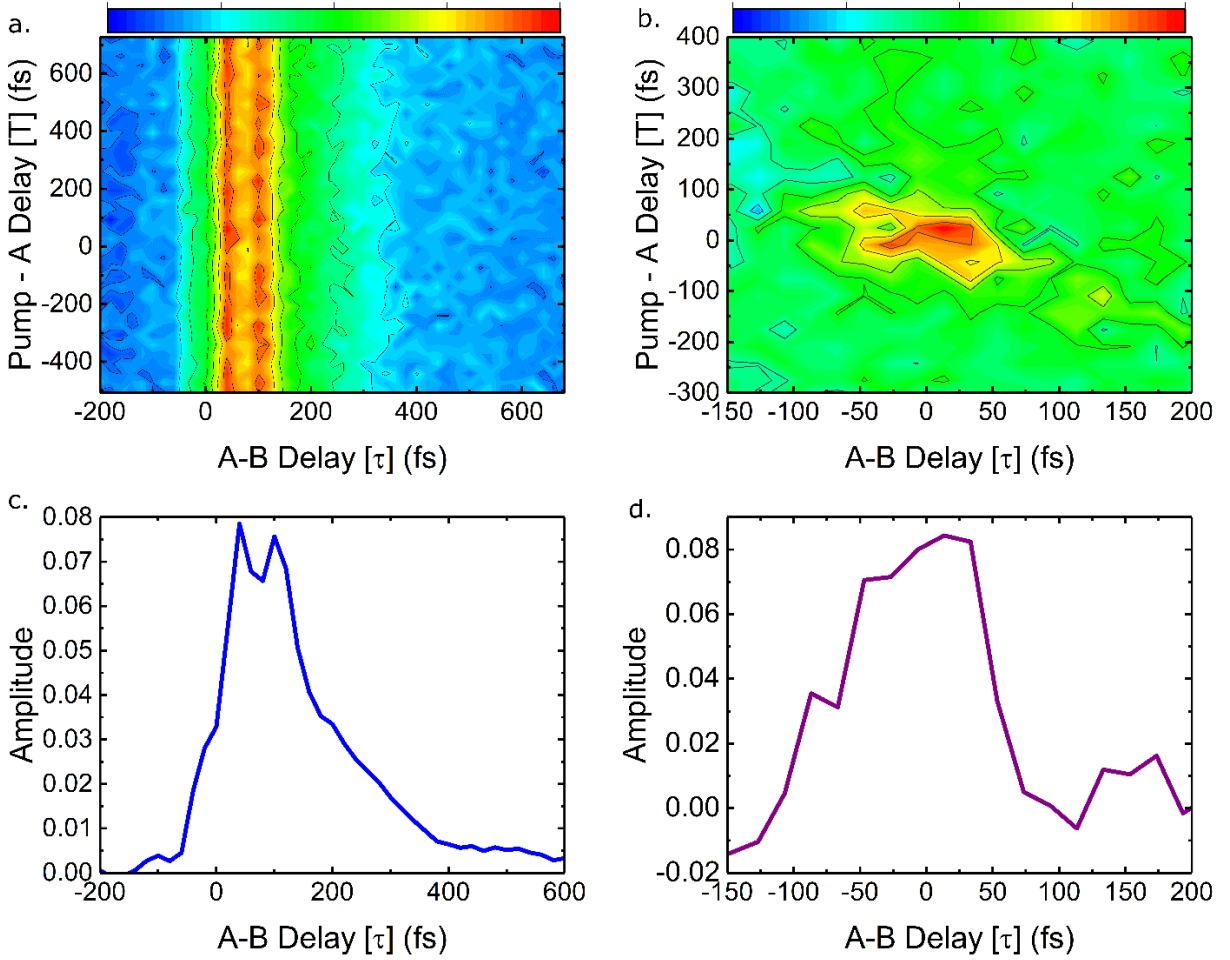

Figure 3: (a) Measured 2D time-resolved homodyne OKE signal without the excitation pump pulse and (b) UTPS signal with the pump pulse. The signal is measured in the direction of probe *B*. Color scales represent signal amplitude in arbitrary units. (c) Projection, and (d) line-out for a pump-probe *A* delays *T*=0, from each 2D plot as a function of *A-B* delay τ. The polarization of the pump and probe *A* are aligned at 45°, while probe *B* polarization is 0°. For positive delays τ > 0, probe *A* arrives before probe *B*.

For isotropic media, $\chi^{(3)}_{eff}$ contains two non-zero tensor elements $\chi^{(3)}_{yxyx}$ and $\chi^{(3)}_{yxxy}$, and, for anisotropic media, two additional terms $\chi^{(3)}_{yxyy}$ and $\chi^{(3)}_{yxxx}$ are also important.[21,23,24] For liquid media, conventional OKE spectroscopy (without an excitation pump pulse) is sensitive to intramolecular vibrational motion and intermolecular librational motion of molecules in the ground electronic state. With the addition of a preceding pump pulse, UTPS allows the coupled electronic and nuclear dynamics of excited electronic states to be probed by OKE spectroscopy. The pump pulse contributes to the measured response function by the addition of excited state 3$^{rd}$-order response terms. The homodyne signal,

$$V_{homodyne}(t) \propto \int_{-\infty}^{\infty} dt'\, I_{probeB}(t-t') \left[ \int_{-\infty}^{\infty} dt''\, R(t'-t'') I_{probeA}(t-t') \right]^2 \quad (2a)$$

$$\mathcal{F}\{R\}^2 \cong \omega^2 \left| \chi_{eff}^{(3)} \right|^2 \quad (2b)$$

is dependent on the square magnitude of the 3rd-order response function $R(t)$, convolved with the probe *A* and *B* pulses.[21,23,24] Each pulse is independently characterized by FROG, so that $R(t)$ can be isolated from the instrument response function (IRF). Since we modify the system with an initial excitation, $R(t)$ contains the excited state dynamics as a function of time delay. This allows for a measurement of $\left|\chi_{eff}^{(3)}\right|$ through $R(t)$ as in Equation 2b.

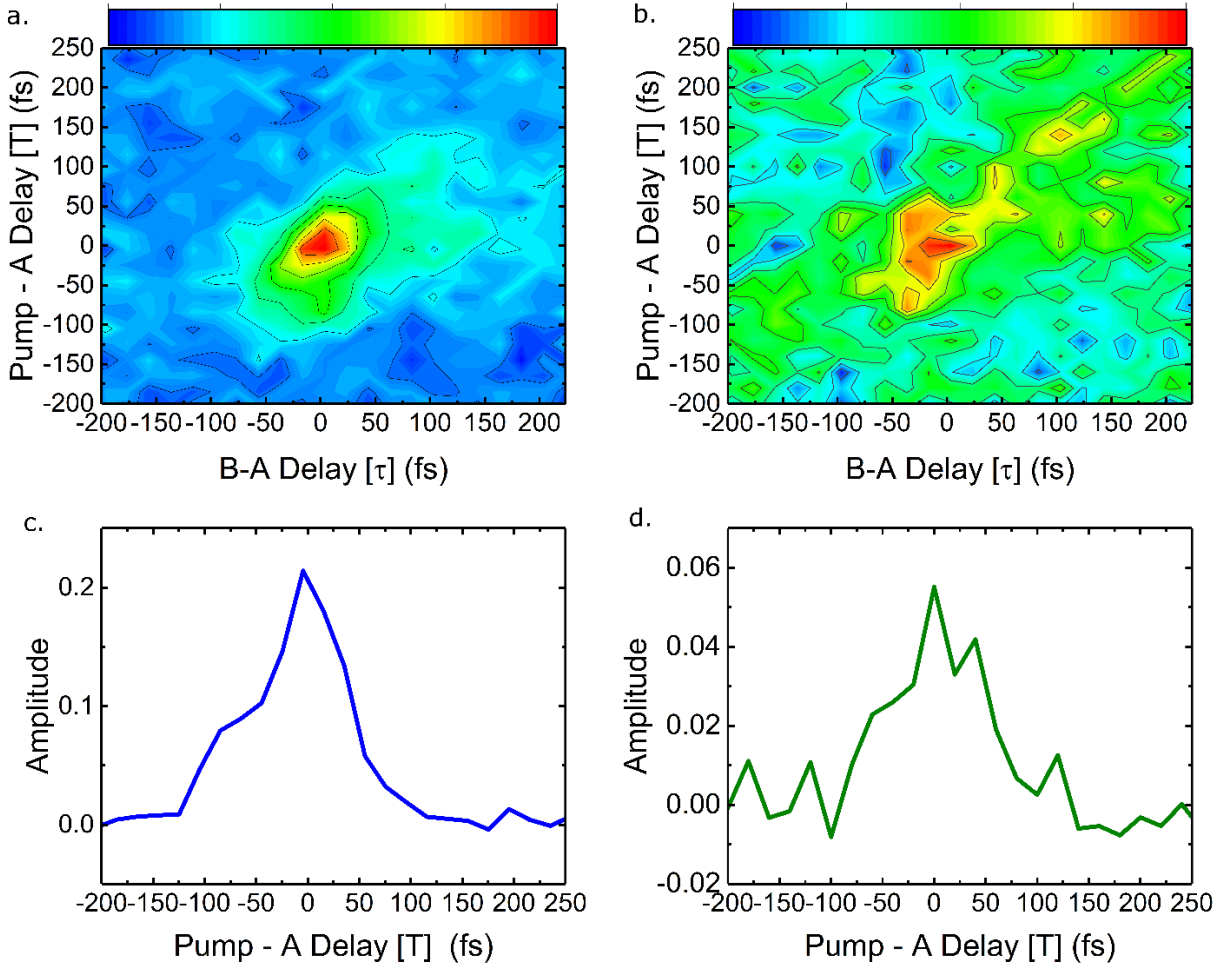

Figure 4: (a) UTPS results for photoexcited nitrobenzene, by optical heterodyning with a local oscillator in-phase with the probe pulse and (b) π/2 out-phase with the probe pulse. The signal is measured in the direction of probe *A* and the pump pulse is linearly polarized parallel to the probe *A* polarization. The color scales represent signal amplitude in arbitrary units. For positive delays *T*, the pump arrives before probe *A*, and for positive delays τ probe *B* arrives before probe *A*. (c) Vertical line-out extracted from the 2D data of panel (a), for *A-B* delay τ=0, showing the pump-probe *A* delay *T* time dependence. (d) Vertical line-out extracted from the 2D data of panel (b), for *A-B* delay τ=0, showing the pump-probe *A* delay *T* time dependence.

Optical heterodyning measurements, using a local oscillator that is in-phase or $\pi/2$ out-of-phase with the probe pulse, produce a signal proportional to the real or imaginary parts, respectively, of $R(t)$.[21,23,24] The local oscillator simplifies the homodyne expressions to produce an OKE signal that is directly proportional to the convolution of the IRF and the 3rd-order response function:

$$V_{heterodyne}(\tau) \propto \int_{-\infty}^{\infty} dt' R(\tau - t') G_0^{(2)}(t') \tag{3a}$$

$$\mathcal{F}\{R(\tau)\} \cong \omega \left[\chi_{eff}^{(3)}\right] \tag{3b}$$

Optical heterodyning measurements therefore allow the real and imaginary components of the effective 3rd-order response to be reconstructed through Fourier analysis,

$$V_{in-phase}(\tau) + i\, V_{out-of-phase}(\tau) \propto V_{heterodyne}(\tau) \tag{4}$$

An example of a time-resolved homodyne OKE measurement in liquid nitrobenzene, measured in the direction of probe pulse *B* (detector PD1 in Figure 1), is presented in Figure 3a. The homodyne OKE signal (Figure 3a) includes the intramolecular vibrational and intermolecular librational dynamics on the ground electronic state of the liquid sample, as well as the purely electronic response to the probe *A* and probe *B* pulses, which is due to the probe-induced polarization[25], at early times within the *A-B* cross correlation of 57 fs (full width at half maximum, FWHM). Figure 3b shows the results of homodyne UTPS experiments upon photoexcitation of nitrobenzene with pump polarization set to 45° relative to probe *B*. The measured UTPS signal exhibits a pump-probe *A T*-dependence that decays over 200 fs, considerably longer than the pump-A-B 3-pulse cross-correlation, which is 69 fs FWHM.

Examples of heterodyne UTPS measurements for photoexcited nitrobenzene are shown in Figures 4a and b, for in-phase and $\pi/2$ out-of-phase local oscillator, respectively. Here the signal is measured in the direction of probe pulse *A* which has polarization parallel to the pump polarization.

Discussion

The homodyne UTPS signal of Figure 3b depends quadratically on the magnitude of $\chi_{eff}^{(3)}$ (Equation 2) and contains information on the coupled electronic and nuclear dynamics of excited electronic states following population of $S_1$ by the pump pulse. The present discussion focuses on the technical aspects of the experimental measurements, therefore we limit the discussion of the dynamics of the photoexcited nitrobenzene sample to a qualitative description. The dynamics of photoexcited electronic states of benzene have been investigated previously in transient grating spectroscopy experiments[26,27], time-resolved photoelectron spectroscopy[28], state-specific fragment imaging[29], and theoretically using high-level electronic structure calculations[30,31]. The shape of the homodyne UTPS signal (Figure 3b) follows a rapid increase near $T$=-50 fs and $\tau$=-50 fs, a symmetric peak near $T$=0 fs and $\tau$=0 fs, consistent with the coherent electronic response within the 69 fs duration of the instrument response function, and an exponential decay with some weaker structures between $\tau$=50 fs and $\tau\approx$300 fs. These timescales are consistent with the decay times measured in previous transient grating spectroscopy experiments,[27] and recent electronic structure calculations,[30,31] that have established a rapid internal conversion due to strong coupling of the first excited singlet electronic state $S_1$ to the triplet manifold, which couples relatively weakly with the ground electronic state. This path competes with a second decay path involving a conical intersection between $S_1$ and the ground electronic state. In comparison, the homodyne OKE signal (Figure 3a), which is sensitive only to dynamics on the ground electronic state, is distinguished by a double-peak structure at A-B delays of 50 to 100 fs, followed by a significantly longer exponential decay, with significant signal remaining after 300 fs.

The heterodyne signal, measured in the direction of probe pulse $A$, exhibits a rapid decay in both pump-probe $A$ delay T and probe $B$-$A$ delay $\tau$. The magnitude of the heterodyne signal is considerably larger for the in-phase local oscillator, compared to the out-of-phase heterodyne measurement, which indicates that the real part of the effective 3rd-order response, corresponding to a nonlinear birefringence, is larger than the imaginary part, which is a measure related to the nonlinear absorption (Equation 4).

This experimental technique offers a particularly strong connection with modern quantum chemistry methods and to model UTPS signals. Because UTPS has a well-described set of tensor elements that contribute to $\chi^{(3)}_{eff}$, many theoretical techniques can be used to understand the resulting signals. Quantum chemistry software packages, such as Dalton, can be used to quantify $\chi^{(3)}_{eff}$ through response theory calculations that allow for the application of electronic structure theory to the calculation of tensor elements of the second order hyperpolarizability in the molecular frame (which is directly analogous to the lab frame bulk $\chi^{(3)}$).[32–35] Additionally, detailed models that describe how particular classes of systems, like those of conjugated molecules, can also be used to describe how $\chi^{(3)}_{eff}$ will evolve.[36,37] By translating model predictions of $\chi^{(3)}$ to $\chi^{(3)}_{eff}$, these methods can be used to predict UTPS signals.

While the UTPS technique presented relies on three NIR pulses to excite and probe molecular dynamics, this technique has potential application with a wide range of photon energies including those in the extreme ultraviolet (XUV) and x-ray energy regime. Such photon energies applied to ultrafast nonlinear spectroscopies bring many new possibilities for real-time measurements of ultrafast processes in molecules, such as nonadiabatic electronic and nuclear dynamics near conical intersections[38], and electronic coherences in excited systems.[39] Resonant core to valence electronic transitions are typically well-localized on a specific site of the molecule, therefore XUV and soft x-ray probes could be employed to probe electronic dynamics at a specific moiety or atom within a molecular system. Recent examples of XUV nonlinear spectroscopy include transient grating measurements in $SiO_2$ by two XUV pulses,[40] using a near-UV probe to detect the four-wave mixing (FWM) signal at near-UV wavelengths in a non-colinear geometry. Ding *et al*[41] recently reported their investigation of coherent nonlinear

XUV absorption in the $Ne^{2+}$ ion, using phase-locked pulses from a free electron laser. Ding *et al.*[42] and Neumark, Leone and coworkers[9,10,43,44] independently realized non-colinear FWM in atomic gases with attosecond XUV and femtosecond NIR pulses. Neumark, Leone and coworkers extended their approach to reveal coupled electronic and nuclear dynamics in molecules[45,46], using near infrared pulses to couple excited valence states in $H_2$ and the bright valence b′ state and the dark a″ double-well state of $N_2$. Employing pulse shaping on one NIR pulse, they have recently demonstrated two-dimensional NIR-XUV FWM spectroscopy[47], which is differential in both the NIR interaction energy and the XUV emission energy. This multidimensional FWM spectroscopy method reveals electronic dynamics and separates different pathways that were shown to be crowded in FWM experiments without NIR pulse shaping. Delivering sufficiently intense coherent XUV pulses to a sample to probe a nonlinear response is a challenge that can be met only if the XUV beamline is efficient, which is usually met by minimizing the number of steering and focusing mirrors. The further development of femtosecond time-resolved nonlinear spectroscopy with XUV photon energies will clearly benefit from techniques that require few noncolinear pulses and simple optical arrangements, which are inherent to UTPS. We anticipate that the new technique described here will enable future applications driving nonlinear processes with intense XUV pulses, to combine the multidimensional capabilities and sensitivity of nonlinear spectroscopies with the atom-specificity.

Conclusions

We present ultrafast transient polarization spectroscopy as a new experimental technique to probe the coupled electronic and nuclear dynamics of excited electronic states in a non-resonant, non-colinear 3-pulse scheme. The UTPS technique enables optical heterodyning or homodyning in the same experimental setup and, since the probe pulses weakly interact with the sample, sample damage by ionization or photolysis is avoided. The orthogonal polarizations of the signal and probe allow the signal emitted by the photoexcited sample to be isolated from the probe beam, removing the probe background that is common with many other ultrafast spectroscopic techniques. The simple optical and optomechanical instrumentation requires only three pulses interacting at the sample, and only two optical delay lines, therefore the technique is readily extendable to tunable resonant or non-resonant excitation pulses from longer wavelengths out to the extreme ultraviolet or soft x-rays.

Acknowledgements

This work was supported by the U.S. Department of Energy, Office of Science, Basic Energy Sciences, Chemical Sciences, Geosciences, and Biosciences Division.

Author Information:

2. Current address: Department of Physics and Astronomy, Purdue University, West Lafayette, IN 47907 USA

Corresponding Author: *Email: dsslaughter@lbl.gov

ORCID :
Daniel S. Slaughter: 0000-0002-4621-4552
Matthew M. Brister: 0000-0002-1563-6496